\documentclass{aa}
\usepackage{natbib} \bibpunct{(}{)}{;}{a}{}{,}
\usepackage{txfonts}
\usepackage{graphicx}

\begin{document}
\title{Simulations of planet-disc interactions using Smoothed Particle
Hydrodynamics}
\author{Christoph Sch\"afer\inst{1} \and Roland Speith\inst{2,
1} 
\and Michael Hipp\inst{3} \and Wilhelm Kley\inst{1}}
\institute{Computational Physics, Auf der Morgenstelle 10C, 72076
T\"ubingen, Germany 
\and Department of Physics and Astronomy,
University of Leicester, University Road, Leicester LE1 7RH, United
Kingdom
\and Wilhelm-Schickard-Institut f\"ur Informatik, Sand 13, 72076 T\"ubingen, Germany}
\date{Received --- / Accepted ---}

\abstract{We have performed Smoothed Particle Hydrodynamics (SPH)
simulations to study the time evolution of one and two protoplanets
embedded in a protoplanetary accretion disc. We investigate accretion
and migration rates of a single protoplanet depending on several
parameters of the protoplanetary disc, mainly viscosity and scale
height. Additionally, we consider the influence of a second
protoplanet in a long time simulation and examine the migration of the
two planets in the disc, especially the growth of eccentricity and
chaotic behaviour. One aim of this work is to establish the
feasibility of SPH for such calculations considering that usually only
grid-based methods are adopted. To resolve shocks and to prevent
particle penetration, we introduce a new approach for an artificial
viscosity, which consists of an additional artificial bulk viscosity
term in the SPH-representation of the Navier-Stokes equation. This
allows accurate treatment of the physical kinematic viscosity to
describe the shear, without the use of artificial shear viscosity.

}
\titlerunning{Simulations of planet-disc interactions using SPH}
\maketitle
% Introduction
\section{Introduction}
Since the discovery of the first extrasolar planet in the mid-nineties of the 
last century by \cite{Mayor:1995:jupiter}, the interest in planet
formation research has grown rapidly. By now about 120 extrasolar
planets are known. For comprehensive reviews on extrasolar planets and their
detection we refer to \cite{2000RPPh...63.1209P} and
\cite{2000prpl.conf.1285M}. 
These newly discovered planetary systems partially feature attributes that differ
enormously from those of our solar system. In particular, orbital
semi-major axes of down to $2.25\times10^{-3}$~AU, masses in the range of $0.12-13.75$~M$_\mathrm{jup}$, 
and large eccentricities of up to $0.927$ have been found, 
in contrast to the planets in the solar system which orbit the sun 
at larger distances on almost circular orbits with rather lower masses. 
Therefore, nowadays planet formation theory not only has to explain the formation of
our solar system, but the formation of planetary systems with these
substantially different features as well. 

It is generally believed that planet formation is part of 
star formation. Stars are formed by gravitational collapse of a
molecular cloud. During this process, angular momentum conservation leads
to the formation of a circumstellar disc around a central object, the
protostar. The material in the disc slowly accretes onto the protostar in
the centre until the protostar becomes a star. Circumstellar discs
composed of
dust and gas have been found around young T Tauri 
stars \citep{1996Natur.383..139B}. Star formation theory suggests a
typical lifetime of such circumstellar discs in the range of $10^6$ to
$10^7$~yr, which determines the time frame for planet formation (see
\cite{2000fdtp.conf....1K} and references therein).
It is assumed that planets form in these circumstellar (or
protoplanetary) discs. Two mechanisms are proposed,  
either fast formation by direct fragmentation due to gravitational instabilities 
\citep{2002Sci...298.1756M, 2001ApJ...563..367B, 2003LPI....34.1075B}, 
or slow formation during a process of several
steps: at first the dust in the disc grows via a so-called hit-and-stick
mechanism and settles in the mid-plane of the disc, where a dense layer
forms \citep{2000prpl.conf..533B}. There the solid material is able to
combine into larger bodies,
which eventually form planetesimals. Finally, planetary-sized objects
develop out of these planetesimals by collisions
\citep{2002Icar..155..436O}. The latter, slower scenario leads to a
longer formation time. Calculations by \cite{1996Icar..124...62P}
suggest a formation time of $1$ to $10$ million years for Jupiter and Saturn, and
of $2$ to $16$ million years for Uranus, respectively.

This scenario leads to essentially circular orbits of the planets
as they form in a keplerian disc. Due to the lack of material in
the inner regions of the disc,
more massive planets can only be created at larger distances to the central
object. However, while these properties fit well to our solar system where the
eccentricities are rather small and the massive planets are located in
the outer regions, the model cannot explain the appearance of
highly massive planets on significantly eccentric orbits at orbital radii down to $0.04$ AU, as found among
the extrasolar planets. 

A possible solution to this problem is planet migration. It is generally assumed
that these planets have formed further out in the disc and have migrated
inwards later. This migration has its cause in 
tidal interactions of the planet with the disc
\citep{1980ApJ...241..425G}. The planet exerts tidal
torques on the disc which lead to the formation of spiral density
wave perturbations. Simultaneously, since the disc loses its
axisymmetry, a net torque acts on the planet itself, which starts
migrating in radial direction \citep{2000prpl.conf.1111L}. 
The migration rate depends on different factors, such as the
mass of the planet, the density distribution in the disc and the
viscosity of the disc material. 

Another problem lies in the high masses of the discovered planets. The tidal
interaction of the planet can eventually lead to the opening of a gap at
the planet's orbital region in the disc, which has a severe impact on the
accretion rate. Until the numerical calculations by
\cite{1996ApJ...467L..77A}, who modelled the accretion on 
circumbinaries, and the simulations of a jovian protoplanet embedded in a disc by 
\cite{Kley:1999:Mass,Bryden:1999:tidally}, it has been assumed that accretion stops after
gap formation. These simulations however yield a final non-vanishing accretion rate
which depends on the viscosity in the disc. Since then, further work has
been done on this by various other authors 
\citep{Lubow:1999:Disk,Bryden:2000:protoplanetary,
2000prpl.conf.1111L,Terquem:2000:disks,Nelson:2000:migration}.

While the numerical method known as Smoothed Particle Hydrody\-namics
(SPH) has been used for one of the first simulations that examined the 
system of an embedded protoplanet in a disc \citep{1987EM&P...39....1S}, 
later simulations were nearly exclusively performed with grid-based methods, mainly because computer resources
have not been available for other high resolution simulations. This has
changed only recently. In this paper
we present a feasibility study for the simulation of planet-disc
interactions with SPH, applying a new approach for the artificial
viscosity, and we show the results of simulations with one
and two embedded protoplanets in a viscous accretion disc.
Very recently, also \cite{2004MNRAS.347..421L}
have used SPH for investigating planet-disc interaction. However,
they focus on the scenario where planets form due to selfgravitating
instabilities, while we consider non-selfgravitating discs 
%% WK
in two dimensions ($r-\varphi$).

The outline of the paper is as follows.
In the next section we present the physical model of the system,
including an overview on gap formation and planet migration theory. Numerical
issues will be discussed in Sect.~\ref{section_numerics}, where especially
the approach for a new artificial bulk viscosity will be described. The
results of several simulations will be given in \ref{section_results},
followed by a discussion in Sect.~\ref{section_discussion}. 
Finally, we draw our conclusions in Sect.~\ref{section_conclusions}. 
\section{Physical model}
We only consider a non-selfgravitating, thin accretion disc model for the
protoplanetary disc. The disc is located in the $z=0$ plane, rotating
around the $z$-axis. The vertical thickness $H$ of the disc is assumed
to be small in comparison to the radial extension $r$, $H/r \ll 1$.
Therefore it is possible to vertically integrate the hydrodynamic
equations and use vertically averaged quantities only,
neglecting the
vertical motion in the disc. The protostar in the centre of the disc
always is assumed to have a mass of $M_\mathrm{c}=1\mathrm{M}_{\sun{}}$.
\subsection{Basic equations}
The equations describing a viscid flow of gas in the disc are the
continuity equation and the Navier-Stokes equation. In the Lagrangian scheme the
former reads
\begin{equation}
\frac{\mathrm{d} \Sigma}{\mathrm{d}t} + \Sigma \vec{\nabla v} = 0, 
\end{equation}
where $\Sigma$ is the surface density and $\vec{v}$ the
velocity. The surface density is defined by
\begin{equation}
\Sigma = \int_{-\infty}^{\infty} \varrho \mathrm{d}z,
\end{equation}
where $\varrho$ denotes the density.
The Navier-Stokes equation in Lagrangian formulation reads
\begin{equation}
\frac{\mathrm{d}\vec{v}}{\mathrm{d}t} = -\frac{1}{\Sigma}  \vec{\nabla} p + \vec{f} + 
\frac{1}{\Sigma} \vec{\nabla} \tens{T},
\end{equation}
where $p$ is the vertically integrated pressure, and $\vec{f}$ denotes a
specific external force, i.e., the gravitational acceleration due to
the star and the $N_\mathrm{p}$ planets with masses $m_i$ and locations $\vec{r_i}$
\begin{equation}
\vec{f} =   -  G M_\mathrm{c} \frac{\vec{r}-\vec{r}_{\mathrm{c}}}{(\vec{r}-\vec{r}_\mathrm{c})^3} 
- \sum_{i=1}^{N_{\mathrm{p}}} G m_i
\frac{\vec{r}-\vec{r}_i}{(\vec{r}-\vec{r}_i)^3}.
\end{equation}
The last term in the equation of motion describes the viscous forces and
contains the viscous stress tensor which can be written as
\begin{eqnarray}
\tens{T}_{\alpha\beta} & = & \eta \left( \frac{\partial v_\beta}{\partial
x_\alpha} + \frac{\partial v_\alpha}{\partial x_\beta} - \frac{2}{3}
\delta_{\alpha\beta} \frac{\partial v_\gamma}{\partial x_\gamma}\right)
+ \zeta \delta_{\alpha \beta} \frac{\partial v_\gamma}{\partial x_\gamma}\\
& \equiv & \eta \sigma_{\alpha\beta} + \zeta \delta_{\alpha \beta} \frac{\partial v_\gamma}{\partial x_\gamma},
\label{viscous_stress_tensor}
\end{eqnarray}
(Greek indices denote spatial coordinates, and Einstein's summing
convention holds throughout),
where the coefficients of shear viscosity, $\eta$, and bulk viscosity, $\zeta$, 
are positive and do not depend on the velocity.
\subsection{Equation of state}
For computational simplicity we follow the usual approach and use an 
isothermal equation of state, where the vertically integrated pressure $p$ is related
to the surface density $\Sigma$ by
\begin{equation}
p = c_\mathrm{s}^2 \Sigma.
\end{equation}
Here the local isothermal sound speed is given by
\begin{equation}
c_\mathrm{s} = \frac{H}{r} v_\mathrm{kep},
\end{equation}
where $v_\mathrm{kep}$ denotes the Keplerian velocity of the unperturbed
disc. The scale height of the disc, which is defined as the ratio of the
vertical height to the radial distance, $H/r$, equals the inverse Mach number
${\cal M}$ of the flow in the disc and is taken as an input parameter 
for our disc model. 
By choosing such an equation of state we suppose that the disc stays
geometrically thin and emits all thermal energy which results
from tidal and viscous dissipation locally.
\subsection{Viscosity}
Viscous processes are of great importance in the theory of accretion
discs. In order to move radially inwards, the disc material
has to lose its angular momentum. Therefore a permanent flow of angular
momentum radially outwards and of mass radially inwards takes place in the disc. 
Often, theories of protostellar discs assume
viscous accretions discs, see e.g.\ \cite{Papaloizou:1999:discs}.
We follow the prescription by \cite{Shakura:1973:Black} and assume an
effective $\alpha$-viscosity, where the kinematic viscosity $\nu$ in the disc
is given by 
\begin{equation}
\nu = \alpha c_\mathrm{s} H. 
\end{equation}
The shear viscosity $\eta$ is related to the kinematic viscosity by
$\eta = \nu\Sigma$. 
Typical $\alpha$-values for protostellar discs are in the range of $10^{-2}$ to $10^{-3}$. 
\subsection{Gap formation and planet migration}
Using perturbation theory, the tidal interaction of the planet and the accretion disc can be
analyzed analytically if the planetary mass is much smaller than the
mass of the central object, see 
\cite{1979ApJ...233..857G, 1980ApJ...241..425G} for details. 
The planet exerts torques on the disc material (and vice versa), which can lead to gap
opening at the planet's orbital region if these tidal torques exceed the
internal viscous torques in the disc \citep{1984ApJ...285..818P}.
The faster (slower) moving disc material inside (outside) the planet's
orbit exerts a positive 
(negative) torque on the planet. The planet will begin to migrate
radially as soon as these torques do not cancel out. There are two
different timescales for migration depending on whether a gap has formed
or not, which are called type I and type II migration respectively
\citep{1997Icar..126..261W}.
\paragraph{Type I migration.}
No gap formation occurs as long as the response of the disc to the tidal
perturbation stays linear. The radial migration velocity of the planet
is then given by \citep{Nelson:2000:migration}
\begin{equation}
\frac{\mathrm{d}r}{\mathrm{d}t} \sim -c_1 q \left( \frac{\Sigma
r^2}{M_\mathrm{c}}\right) \left( r \Omega_\mathrm{kep} \right) \left( \frac{r}{H}
\right)^3,
\end{equation}
where $q$ denotes the mass ratio of the planet to the central star,
$M_\mathrm{c}$ the mass of the central star, and $\Omega_\mathrm{kep}$
the Keplerian angular velocity. The factor $c_1$ accounts for the
imbalance of the torques between the two regions of the disc inside and
outside the planet's orbit. Hence, the radial migration velocity will 
be larger for a more massive planet as long as the response of the 
disc stays linear. 

As soon as the planet has
grown to a sufficient high mass, the response of the disc to the
perturbation becomes non-linear, and gap formation occurs. 
\paragraph{Type II migration.}
In order to form a gap the tidal torques exerted by the planet have to
exceed the internal viscous torques in the disc. The conditions for gap
formation have been widely studied in detail by \cite{1984ApJ...285..818P},
\cite{1985prpl.conf..981L,1993prpl.conf..749L}, \cite{Bryden:1999:tidally}
and \cite{Nelson:2000:migration}. Here we only present the different
truncation conditions and scenarios for type II migration.
A sufficient condition
for viscous truncation is 
\begin{equation}
q > \frac{40\nu}{\omega r^2} \label{viscous_truncation},
\end{equation}
where $\omega$ is the angular velocity of the planet and $r$ its orbital
radius. For discs with a very low viscosity \cite{1996ApJS..105..181K}
suggest the thermal truncation condition
\begin{equation}
q > \left( \frac{H}{r} \right)^3. \label{truncation_condition}
\end{equation}
During type II migration, the migration of the planet is driven by the
viscous evolution of the disc. The migration rate in this case is given
by the radial drift velocity of the gas in the disc due to viscous
processes \citep{1984ApJ...285..818P}
\begin{equation}
\frac{\mathrm{d} r}{\mathrm{d}t} \sim \frac{3\nu}{2r}.
\label{equation_visc}
\end{equation}
This relation holds as long as the mass of the planet is lower than the
mass of the disc material with which it interacts. Otherwise the inertia
of the planet gets important and slows down the orbital evolution. This
has been studied in detail by \cite{1995MNRAS.277..758S} and
\cite{1999MNRAS.307...79I}. \cite{Nelson:2000:migration} find a
migration rate based on a model by 
\cite{1999MNRAS.307...79I} which reads
\begin{equation}
\frac{\mathrm{d} r}{\mathrm{d}t} \sim \frac{3\nu}{2} 
\left( 10\pi^4\Sigma^4\right)^{1/5} m_\mathrm{p}^{-4/5}
r^{3/5},
\label{inertia_migration}
\end{equation}
where $m_\mathrm{p}$ is the mass of the planet.
% Numerics
\section{Numerical issues\label{section_numerics}}
We apply Smoothed Particle Hydrodynamics (SPH) for our simulations. SPH is a grid-free Lagrangian particle
method for solving the system of hydrodynamic equations for compressible
and viscous fluids first introduced by
\cite{Gingold:1977:SPH} and \cite{Lucy:1977:numerical}. SPH is
especially suited for boundary-free problems with high density contrasts. 
Rather than being solved on a grid, the equations are solved at the positions 
of the so-called particles, each of which represents a volume of the
fluid with its physical quantities like 
mass, density, temperature, etc.\ and moves with the flow according to the 
equations of motion. For a more detailed description see, e.g.,
\cite{1990nmns.work..269B} and \cite{Monaghan:1992:Sph}. 
\subsection{The code}
Our SPH-code is fully parallelized to make optimal use of
today's supercomputers which is necessary to achieve the required
resolution and accuracy. 
The parallel implementation is done with the so called
ParaSPH Library, a generalized set of routines to simplify and
optimize the development of parallel particle codes \citep{bubeck98parallel}.
This library allows a clear separation between the
implementation of the physical problem to be simulated and the parallelization. A programmer
using the library works with parallel arrays containing the physical
quantities of the SPH-particles and iterators to step through the
particles and their neighbours.  Domain decomposition, load balancing,
nearest neighbour search and communication is hidden in the library.
ParaSPH uses MPI (Message Passing Interface) for the communication and works on
every parallel architecture providing an MPI library. On a Cray T3E one
can achieve a parallel speedup of about 120 using 256 nodes for a problem
with 360\,000 particles. On the Beowulf-Cluster where we have performed 
the majority of our simulations -- a Pentium III based
Linux-Cluster with a Myrinet communication network located at the
University of T\"ubingen -- the parallel speedup is about 60 on 128 processors.
\subsection{Physical and artificial viscosity, XSPH}
To model the physical shear viscosity we use the implementation
introduced by \cite{1994ApJ...431..754F} 
and solve the Navier-Stokes equation including the entire viscous
stress tensor (\ref{viscous_stress_tensor}), in contrast to the usual
approach of a scalar artificial viscosity term. For the physical shear
flow, we assume a constant kinematic viscosity $\nu$ and a vanishing
bulk viscosity $\zeta=0$. Then, the acceleration due to the shear
stress tensor $\sigma_{\alpha\beta}$ reads
\begin{equation}
\left. \frac{\mathrm{d}v_{\alpha}}{\mathrm{d}t} \right|_{\mbox{shear}}= 
\frac{1}{\Sigma} \frac{\partial
(\nu\Sigma\sigma_{\alpha\beta})}{\partial x_\beta}.
\end{equation}
Applying the smoothing discretization scheme to the derivative
of the stress tensor yields the SPH representation
\begin{equation}\label{physical_viscosity}
\left.\frac{\mathrm{d}v_{\alpha i}}{\mathrm{d}t}
\right|_{\mbox{shear}} = 
\sum_{j} \frac{m_{j}}{\Sigma_i\Sigma_j}
\left(\nu_i\Sigma_i\sigma_{\alpha\beta i} + \nu_j\Sigma_j\sigma_{\alpha\beta j}\right)
\frac{\partial W_{ij}}{\partial x_{\beta}}
\end{equation}
with the particle form of the shear tensor
\begin{equation}\label{visc2}
\sigma_{\alpha \beta i} =
V_{\alpha \beta i} + V_{\beta \alpha i} - \frac{2}{3}
\delta_{\alpha \beta} V_{\gamma \gamma i},
\end{equation}
where $V_{\alpha \beta i}$ is the particle representation of the velocity
gradient $\partial v_\alpha /\partial x_\beta$
\begin{equation}\label{visc3} 
V_{\alpha \beta i} = \frac{\partial v_{\alpha i}}{\partial
x_{\beta}} =
\sum_{j} \frac{m_{j}}{\Sigma_{j}}
(v_{\alpha j}-v_{\alpha i}) 
\frac{\partial W_{ij}}{\partial x_\beta}.
\end{equation}
The indices $i$ and $j$ denote quantities that are assigned to SPH
particles with positions $\vec{r}_i$ and $\vec{r}_j$, respectively;
$m_j$ is the particle mass, and $W$ is the usual SPH kernel function.
Replacing surface density $\Sigma$ by (volume) density $\varrho$, this
SPH-representation of the shear viscosity also holds in 3D.

In order to prevent the particles from mutual penetration, 
we add an additional artificial bulk viscosity to the viscous stress
tensor. The bulk part of the viscous stress tensor is given by
$\zeta \vec{\nabla}\vec{v}$ and leads to an additional acceleration of particle
$i$ according to
\begin{equation}\label{artbulkvisceq}
\left.\frac{\mathrm{d} \vec{v_i}}{\mathrm{d}t} \right|_{\mbox{bulk}}
= \sum_{j} m_j \zeta_{ij} \frac{(\vec{\nabla} \vec{v} )_i +
(\vec{\nabla} \vec{v})_j}{\Sigma_i\Sigma_j} \vec{\nabla}W_{ij}.
\end{equation}
The artificial bulk viscosity coefficient for the interaction of particles
$i$ and $j$ is given by
\begin{equation}
\zeta_{ij} = \left\{ \begin{array}{ll}
-f \overline{h}_{ij}^2 (\overline{\vec{\nabla} \vec{v}})_{ij} \overline{\Sigma}_{ij}
& \quad  \mbox{for } \vec{v}_{ij}\vec{r}_{ij} < 0 \\
 0 & \quad \mbox{else}
\end{array} \right. ,
\end{equation}
where $h_i$ ($h_j$) is the smoothing length of particle $i$ ($j$),
$\overline{Q}_{ij}$
is an abbreviation for the average $(Q_i + Q_j)/2$ for any quantity $Q$,
and
the notation $\vec{v}_{ij}\vec{r}_{ij} = (\vec{v}_i -
\vec{v}_j)\cdot(\vec{r}_i - \vec{r}_j)$ has been used.
Only approaching
particles are contributing to the artificial viscosity.
The factor $f$ is of order unity and set to $0.5$ in our
simulations.
The divergence of the velocity is calculated according to
\begin{equation}
(\vec{\nabla} \vec{v})_i = \frac{1}{\Sigma_i} \sum_{j} m_j
(\vec{v}_i -\vec{v}_j) \vec{\nabla}W_{ij}.
\end{equation}

To avoid mutual particle penetration, we additionally use the
XSPH-algorithm. The particles are moved at averaged velocity of their 
interaction partners \citep{monaghan:1989:penetration}
\begin{equation}\label{xspheq}
\frac{\mathrm{d}\vec{r_i}}{\mathrm{d}t} = \vec{v}_i + x \sum_{j} \frac{2m_j}{\Sigma_i+\Sigma_j} (\vec{v}_i - \vec{v}_j) W_{ij},
\end{equation}
where $x$ is in the range between 0 and 1.
The use of this averaged velocity keeps the particles in their relative
local order. 

%% RS
In appendix~\ref{appendix_numerics} we demonstrate the low intrinsic
numerical shear stress of this new artificial bulk viscosity approach.
%% RS
\subsection{Migration}
To study the migration of the protoplanet in detail, the
equation of motion of the protoplanet is integrated. The planet's
acceleration due to the gravitational forces reads
\begin{equation}
\frac{\mathrm{d}\vec{v}_\mathrm{p}}{\mathrm{d}t} = GM_\mathrm{c} \frac{\vec{r}_\mathrm{p}
- \vec{r}_\mathrm{c}}{\left| \vec{r}_\mathrm{p} - \vec{r}_\mathrm{c} \right|^3} + \sum_{i=1}^{N}
  Gm_i 
\frac{\vec{r}_\mathrm{p} - \vec{r}_i}{\left(
  \left| \vec{r}_\mathrm{p} - \vec{r}_i\right|^2 + \delta^2 \right)^{3/2}}
\label{grav}
\end{equation}
where $M_\mathrm{c}$ is the mass of the central object, $\vec{r_\mathrm{c}}$ its
position, $N$ the total number of SPH particles in the simulation with
positions $\vec{r_i}$, and $G$ the gravitational constant. 
Only particles which are not inside the Roche lobe of the planet
contribute to the sum. Additionally, the gravitational forces of the
particles on the planet are softened by the use of $\delta^2$, which is
small.  The Roche lobe is approximated using
the formula by \cite{1983ApJ...268..368E}
\begin{equation}
r_{\mathrm{roche}} = \frac{0.49 q^{2/3}}{0.6q^{2/3} + \ln (1+q^{1/3})}
r_{\mathrm{p}}.
\end{equation}
In the case of more than one planet, the additional gravitational forces
of the other objects are added to Eq.~(\ref{grav}), nevertheless this approximation
formula for the Roche lobe is used.
\subsection{Initial conditions}
The initial setup of our standard model places a jovian planet with
mass $m_\mathrm{p} = 1$ M$_\mathrm{jup}$ at
$r_\mathrm{p} = 5.2$ AU and a central object with mass
$M_\mathrm{c} = 1$ M$_{\sun{}}$, which gives a mass ratio of $q= 9.55\times 10^{-4}$.
The surface density $\Sigma$ falls off as $r^{-3/4}$, and the particles
with equal masses are distributed according to this density between
$r_\mathrm{min}=1$ AU and $r_\mathrm{max}=10$ AU. The mass of the disc
is set to $10^{-2}$ $\mathrm{M_{\sun{}}}$ and the initial particle
velocities are $v_r = 0$ and $v_\varphi = v_\mathrm{kep} =
\sqrt{GM_\mathrm{c}/r_i}$ as in the
unperturbed case. The typical number of particles at the beginning of
the simulation is $3\times10^5$. The standard model uses a Mach number
of ${\cal M} = 20$, that is a scale height $H/r = 0.05$, and the kinematic viscosity parameter $\nu_0$ is set
to $10^{15}$ $\mathrm{cm^2/s}$, which corresponds to an $\alpha$ value
of about $4\times10^{-3}$ at the orbital radius of the planet. \par
As the density falls off with radius, fewer particles are located in
the outer regions of the disc than near the central object.
Hence, a constant number of interaction partners is more suited to the 
problem than a fixed smoothing length. Unless otherwise mentioned 
the number of interaction partners per particle is set to $100$. As
mentioned above, we use the XSPH-method to avoid mutual particle
penetration and we set $x=0.5$ in all our simulations.
% Results
\section{Results\label{section_results}}
In this section we present the results of the simulations
with various input parameters. First we consider the case of a single
planet in the disc. Afterwards, the case of two planets and their
mutual interactions is investigated. 
\subsection{One planet}
\subsubsection{Gap formation}
In order to study the formation of a gap, we have performed several
simulations of a disc with an embedded protoplanet. 
The disc scale height and the kinematic viscosity have been used as
input parameters and have been varied accordingly.
Here, the gravitational forces of the
particles on the planet have not been considered, keeping it on a
circular orbit around the central star. Moreover, the planet was
assumed to be able to accrete material without growing in mass, its
Roche lobe has been kept constant throughout the simulations.\par
\begin{figure}
\resizebox{\hsize}{!}{\includegraphics{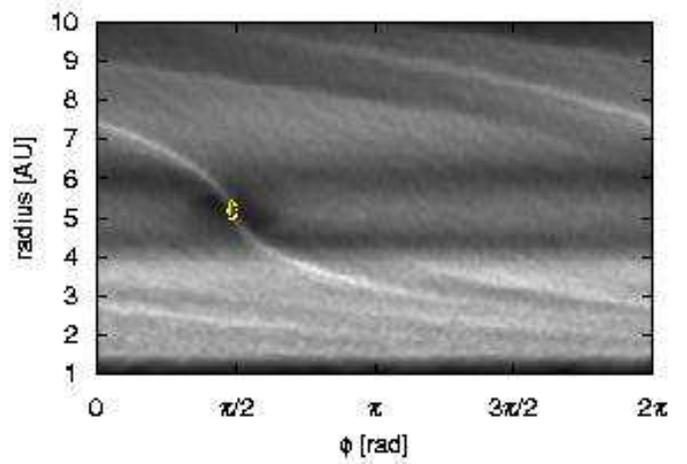}}
\caption{Greyscale plot of the surface density in the disc in the
case of the reference model after 10 orbits of the protoplanet in the
disc. The gap clearing has already started, and the spiral density wave
pattern is visible throughout the disc.
The circle represents the
Roche lobe of the protoplanet according to Eggleton's formula.}
\label{spirals1.eps}
\end{figure}
\begin{figure}
\resizebox{\hsize}{!}{\includegraphics{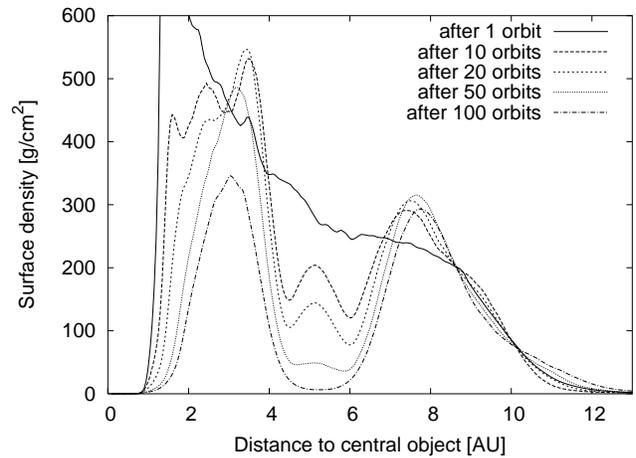}}
\caption{Evolution of the azimuthally averaged surface density for five
different times in the case of the reference model. The planet
has a fixed orbit at $5.2$~AU.}
\label{dens1.eps}
\end{figure}

Just after the start of the simulation the planet disturbs the density
distribution in the disc effectively, which leads to the typical spiral
density wave pattern in the disc which can be seen 
in Fig.~\ref{spirals1.eps}, where a greyscale plot of the surface
density is presented for the system after 10 orbits of the protoplanet. 
In the case of this reference model, gap formation begins
immediately after the start of the simulation run. 
After ten
orbits of the planet the decrease of the density at the protoplanet's
orbital region is already clearly visible. The evolution of the azimuthally
averaged surface density
during the gap formation process is shown in Fig.~\ref{dens1.eps}. 
After one hundred orbits of the protoplanet, the decrease of density at
the orbital region is more than two orders in magnitude. 

\begin{figure}
\resizebox{\hsize}{!}{\includegraphics{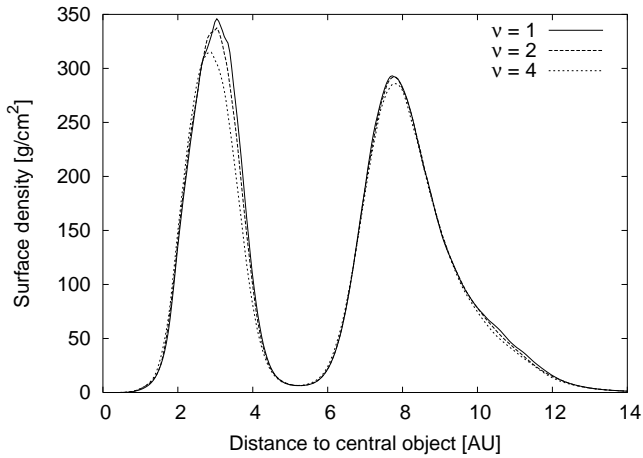}}
\caption{Influence of the kinematic viscosity parameter on the gap size.
The azimuthally averaged surface density is plotted for three different
viscosities in units of $\nu_0=10^{15}$~$\mathrm{cm^2/s}$ after 100 orbits of the protoplanet.
The planet has a fixed orbit with a radial distance of $5.2$~AU.}
\label{visc1.eps}
\end{figure}

\begin{figure}
\resizebox{\hsize}{!}{\includegraphics{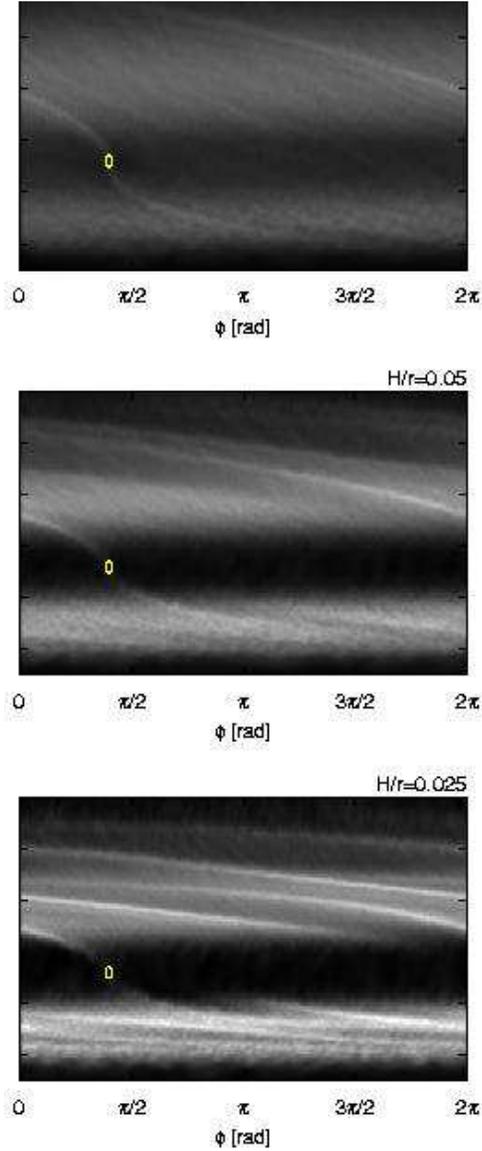}}
\caption{Influence of the scale height of the disc.  The greyscale plot shows 
the surface density in the disc after 100 orbits of the protoplanet for
three different scale heights.}
\label{sh1.eps}
\end{figure}

The radial size of the
gap depends only slightly on the viscosity, as can be seen in Fig.~\ref{visc1.eps}, 
where the surface density is plotted for three simulation runs with
different kinematic viscosity: $\nu = \nu_0 = 10^{15}$~$\mathrm{cm^2/s}$ (reference model), 
$\nu = 2\nu_0$, and $\nu = 4\nu_0$.
A higher viscosity leads to a faster radial motion
of the gas in the disc, and thus to more loss at the
inner boundary, which results in a slightly lower density profile after
the same number of orbits. 
%% RS
A similar behaviour was found by \cite{Kley:1999:Mass}. Because in
Fig.~\ref{visc1.eps} the simulation time is rather short compared with
the viscous timescales, the gap structure looks very similar in each
case. The differences may become more pronounced for longer simulation
runs.  
Unlike the viscosity, the scale height has a severe impact on the depth of 
the gap as can be seen in Fig.~\ref{sh1.eps}. There, the surface
density of the disc after 100 orbits of the protoplanet is plotted for
the scale heights $H/r = 0.1$, $H/r = 0.05$, and $H/r = 0.025$.
The smaller the scale
height, the lower the temperature and the deeper the gap, as can
be seen clearly. Additionally,
the spiral density wave pattern gets tighter as the sound speed is
lower for smaller scale heights.
\subsubsection{Migration and accretion}
In order to study the migration of the protoplanet we consider
again the standard reference model of a jovian protoplanet in a
circular orbit with radius $5.2$~$\mathrm{AU}$.  To achieve a steady-state
particle distribution we proceed as in the last section and let the system evolve 
until the gap is nearly fully developed before we allow the gravitational 
force of the particles to act on the protoplanet.  Then the simulation is restarted and 
the trajectory of the planet is integrated as well.  For the reference 
model we expect pure type II migration on a viscous timescale. 
For this simulation we used nearly 360\,000 particles to model a disc
with mass $3.33 \times 10^{-3}$~M$_{\sun{}}$ and 
$r_\mathrm{in}=1$~$\mathrm{AU}$, $r_\mathrm{out}=13\ \mathrm{AU}$.  
Particles that get closer to the protoplanet than 50\,\% of its Roche
radius are taken out of the simulation and assumed to be accreted by the
protoplanet.  
%% CS
In one model, the mass of the accreted particles is added
to the mass of the planet, in a second, the mass of the planet is
fixed for all times.  The total simulation time is about 30\,000 years.

\begin{figure}
\resizebox{\hsize}{!}{\includegraphics{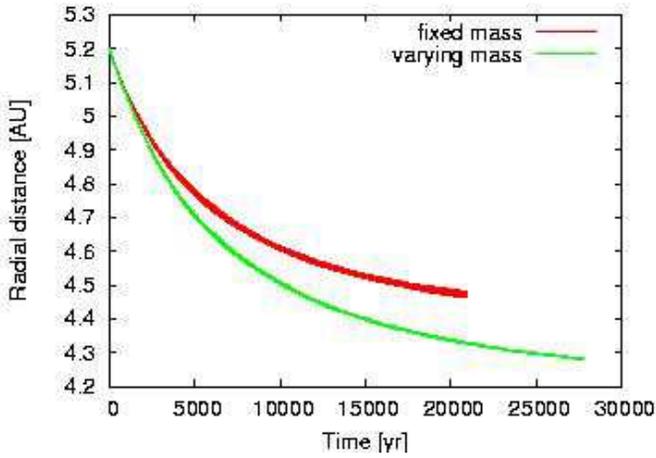}}
\caption{Evolution of the radial distance of the planet to the central
object. The planet is initially located at a radial distance of
$5.2$~$\mathrm{AU}$.  In one model, the accreted mass is added to the
planet's mass, in the other the mass of the planet remains fixed.}
\label{mig1.eps}
\end{figure}

The evolution of the radial distances of the planets is shown in
Fig.~\ref{mig1.eps}.  Initially the migration rate is very high and
about the same for both the planet with fixed mass and the planet with
increasing mass.
After 3\,000 years the radial distances to the central object differ 
distinctly, with the more massive planet migrating slightly faster.
Similar behaviour was found in numerical simulations by \cite{Nelson:2000:migration}.
The planet with a fixed mass ends up at an orbital distance of
$4.5$~$\mathrm{AU}$ after 20\,000~years, which would correspond to a
steady-state migration rate of $3.5 \times 10^{-5}\ \mathrm{AU/yr}$.
The planet with increased mass has a radial distance of
$4.27 \ \mathrm{AU}$ after 
30\,900~years, or a steady-state migration
rate of $1.4 \times 10^{-4}\ \mathrm{AU/yr}$.  As can be seen in
Fig.~\ref{mig1.eps}, the migration rates are not constant but decrease
considerably. This is caused primarily by the mass loss
of the disc, leading to less gravitational interaction between the
planet and the disc.

\begin{figure}
\resizebox{\hsize}{!}{\includegraphics{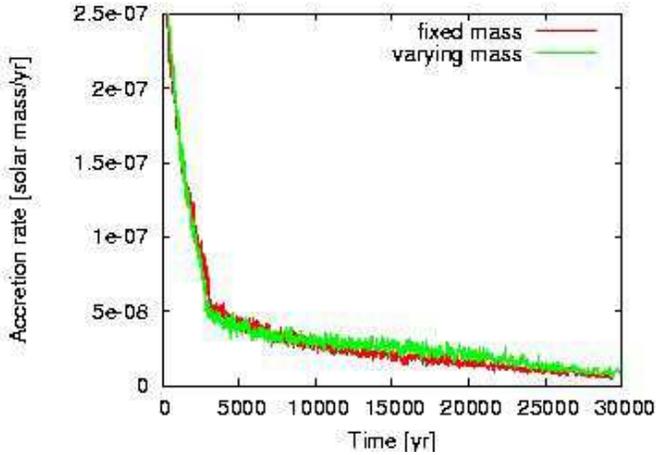}}
\caption{Evolution of the accretion rate on the planet for a planet with
fixed mass and a planet with increasing mass.  The planet is originally
located at a radial distance of $5.2$~$\mathrm{AU}$, and migrates
inwards as shown in Fig.~\ref{mig1.eps}.}
\label{accreted1.eps}
\end{figure}

The accretion rates of the planets for both models are presented in 
Fig.~\ref{accreted1.eps}.  The initial accretion rate of about
$2.5\times10^{-4}$~M$_\mathrm{jup}/\mathrm{yr}$ drops rapidly in the
first 
3\,000~years to a value of $5\times10^{-5}$~M$_\mathrm{jup}/\mathrm{yr}$ for
both the planet with constant mass and the planet with increasing mass.  At
that time, the inner part of the accretion disc has vanished, as the
material was either accreted by the protoplanet or by the central object.
After the loss of the inner disc, the accretion rate decreases more
slowly. Between 3\,000 and 7\,500 years of simulation time, the planet with
increasing mass has a slightly lower accretion rate. The situation
changes after 10\,000 years, when the planet with the fixed mass
accretes less.  At the end of the simulation, both planets have
about the same accretion rate of $10^{-5}$~M$_\mathrm{jup}/\mathrm{yr}$,
although the planet with mass accumulation has already doubled its mass.    
However, during that time the constant loss of material through the open
boundaries and the accretion on the planet lead to a decrease in disc 
mass of nearly one magnitude, which has to be considered for comparisons to 
grid-based simulations with closed boundaries.
% migration
\subsection{Two planets}
As it is generally assumed that the mutual interaction of gravitating objects in an 
accretion disc can finally lead to eccentric orbits, we have included another
protoplanet into our simulations.
\subsubsection{Gap formation} 
\begin{figure}
\resizebox{\hsize}{!}{\includegraphics{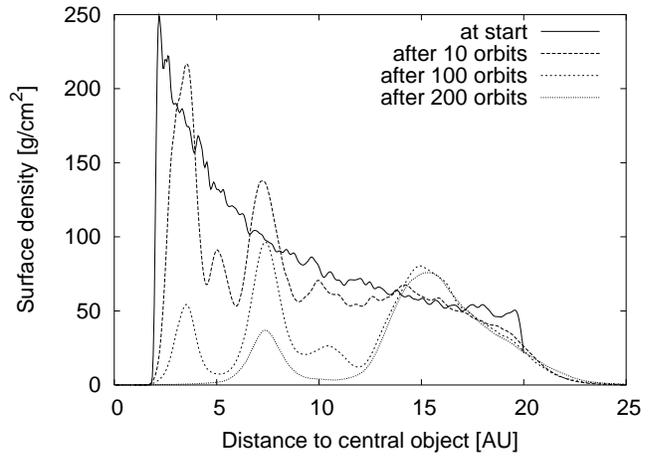}}
\caption{Evolution of the azimuthally averaged surface density in the
case of two protoplanets in the disc. The number of orbits of the inner
planet is indicated. The inner planet is located at $5.2 \
\mathrm{AU}$, the outer planet at twice that distance from the central
object.}
\label{dens2.eps}
\end{figure}

For the simulation with two protoplanets the set\-up was 
chang\-ed as follows: one protoplanet is located at a distance of $5.2 \ \mathrm{AU}$ from
the central gravitating object, the other one at twice that. The
kinematic viscosity parameter is again set to $10^{15}$~$\mathrm{cm^2/s}$, 
and the disc scale height is $0.05$. Now the outer radius of
the disc is $20\ \mathrm{AU}$, and the inner boundary is set to $2
\mathrm{AU}$. The
number of particles in this run was $320\,000$ and the disc mass was set
to $10^{-2}$~M$_{\sun{}}$. The evolution of the density in the disc is 
shown in Fig.~\ref{dens2.eps}. As in the case of a single jovian
planet, gap formation starts immediately after the beginning of the
simulation. This happens for each planet individually, until the particles between the
inner and outer planet have been accreted by the inner planet, and a common
gap has formed. The innermost region of the disc has already vanished after 200~orbits of
the inner planet.
\subsubsection{Migration and accretion}
\begin{figure}
\resizebox{\hsize}{!}{\includegraphics{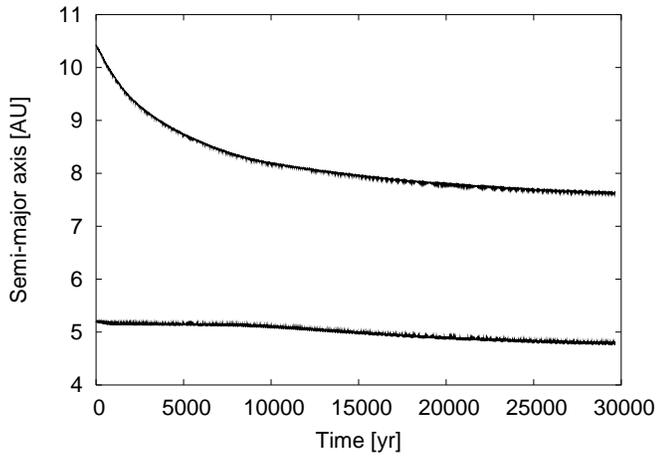}}
\caption{Evolution of the semi-major axis of the two protoplanets. The
initial parameters of the planets are the same as in Fig.~\ref{dens2.eps}, but now
the trajectories of the planets are integrated}.
\label{sma2.eps}
\end{figure}

\begin{figure}
\resizebox{\hsize}{!}{\includegraphics{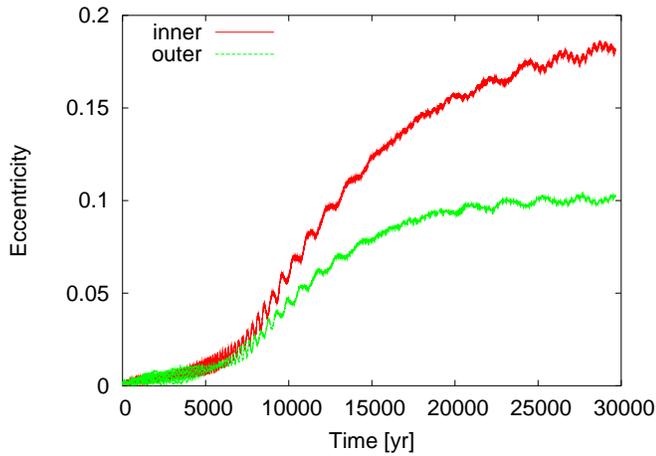}}
\caption{Evolution of the eccentricities of the two protoplanets during
the migration, same simulation as presented in Fig.~\ref{sma2.eps}}.
\label{ecc2.eps}
\end{figure}

After 200~orbits of the inner planet, the simulation is restarted and the
trajectories of the two planets are integrated to examine their migration.
The evolution of the semi-major axis of the two planets is
shown in Fig.~\ref{sma2.eps} for the next 30\,000~years. 
During the first 5\,000~years the inner planet mainly stays on its orbital
radius around the central star because it is not 
surrounded anymore by disc material (see Fig.~\ref{dens2.eps})
which could exert torques. On the other hand
the outer planet migrates inwards due to the action of the outer disc,
resulting in a radial approach of the two planets. As the orbital distance
of the two planets decreases, they capture each other in a 2:1 mean motion
resonance at approximately $t= 7\,500$~years. Upon capture
their eccentricities slowly increase, as
can be seen in Fig.~\ref{ecc2.eps}. 
By the end of the simulation, the eccentricity of the outer planet has
grown to $0.1$ and the eccentricity of the inner planet to $0.18$. In the end,
the planets have reached masses of $1.34$ (inner planet) 
and $2.9$~M$_\mathrm{jup}$ (outer planet), respectively. 
The mass of the disc changed from $10^{-2}$~M$_{\sun{}}$ at the start
of the simulation to $5.64\times10^{-3}$~M$_{\sun{}}$ at
the end.  After all, only 20\,\% of the mass lost by the disc 
have been accreted by the protoplanets. 
%% WK
The resonant capture and
overall evolution of this system
is nearly identical to that described in \cite{2000MNRAS.313L..47K}.
% discussion
\section{Discussion\label{section_discussion}}
In the case of a single jovian planet we obtain the same influence of the
disc scale height on the gap formation and on the structure of the
spiral wave pattern as found already by several other authors (see, e.g.,
\citealt{Bryden:1999:tidally}; \citealt{Kley:1999:Mass}), 
and as postulated by the theory. For the parameters of the reference
model, Eq.~(\ref{equation_visc}) yields a migration rate of $4.13\times10^{-5} \
\mathrm{AU/yr}$ for pure type II migration,
%% CS
for a planet located at a
radial distance of $5.2\ \mathrm{AU}$. We find migration rates of the
same order of magnitude by averaging over the complete simulation time.
The migration rate for this simulation agrees well with 
migration rates found by \cite{2003ApJ...586..540D}, who examined the
migration and growth of one protoplanet in a disc for various initial planet
masses, using a grid-based method and closed boundary conditions.

Interestingly, the accretion rate for a planet with a fixed mass and a
planet with increasing mass do not differ considerably at the end of a
simulation time of 30\,000~years.  Presumably due to the loss of disc
material at the open boundaries, the resolution is too coarse to resolve
the difference correctly.  After a simulation time of 15\,000~years, the
accretion rate onto the planets is about $2.5\times10^{-5} \
\mathrm{M_{jup}/yr}$ for our reference model with a kinematic viscosity
coefficient of $10^{15} \ \mathrm{cm^2/s}$.  This value agrees well with
the results from simulations performed by \cite{Lubow:1999:Disk}, who used the {\tt ZEUS}
hydrodynamics code to simulate disc accretion onto high-mass planets. They
find an accretion rate of $2.13\times10^{-5}\
\mathrm{M_{jup}/yr}$, which is half the value that was published by
\cite{Kley:1999:Mass}.  However, \cite{Lubow:1999:Disk} use different
boundary conditions, namely reflecting closed boundaries at the inner
edge of the disc, and open boundaries at the outer edge.  The
results of the simulations by \cite{Bryden:1999:tidally} provide
accretion rates in the same range, although they use an initial density
profile which is constant throughout the disc.  Extensive simulations
with three different grid codes, {\tt nirvana}, {\tt rh2d}, and
{\tt fargo}, by \cite{Nelson:2000:migration} provide comparable accretion
rates for a jovian planet.

For a single planet in the accretion disc the orbit remains
circular, although the planet migrates towards the central star. This
suggests that the gravitational interaction between the
planet and the protoplanetary disc cannot excite a change in the
eccentricity of the orbit of the planet, at least for low mass 
planets \citep{2001A&A...366..263P}. 
Therefore, in systems with a
single extra-solar planet mostly circular orbits should be expected. This
clearly appears to be in contradiction to the observations, and is why
different mechanisms, e.g.\ resonances, for the
evolution of high eccentricities in single-planet
systems have to be assumed.

We have found that a number of $360\,000$~SPH-particles, which has been
used in the simulation of migration and planet accretion, is
sufficient in 2D to resolve an expected accretion rate of 
several $10^{-5} \ \mathrm{M_{jup}/yr}$. However, the use of an artificial bulk
viscosity is essential, as extensive test calculations have
shown. Without the use of this additional viscosity, the particle orbits
penetrate each other, density shock waves cannot be resolved, and even the
formation of the gap may be suppressed. 
Note that in contrast to the normally used artificial
$\alpha$-viscosity approach (see \cite{Monaghan:1983:shock}) which leads to an
effective physical shear viscosity \citep{1998ApJ...502..342N}, this
artificial bulk viscosity has very little impact on the shear kinematic viscosity in
the disc,
%% RS
as can be seen in appendix~\ref{appendix_numerics}.
%% RS

The number of SPH-particles was not high enough though to  
study the flow around the protoplanet within its Hill radius, as was
done by \citet{2002A&A...385..647D} using nested-grid techniques. This
would have made it possible to examine the accretion process onto the protoplanet
more precisely. \citet{2002A&A...385..647D} find the formation of a
so-called proto-jovian disc around the protoplanet which dominates the
accretion process onto the planet. However, in our simulations the
spatial resolution was 
too low to resolve the formation of such a disc.

%% RS
Generally, SPH seems to be slightly more computationally expensive
for this kind of application than comparable simulations with
grid-based schemes. This is mostly due to the intrinsic noise of the
particle method that requires a comparatively high particle number to
achieve a similar spatial resolution. The advantage of this numerical
method is its Lagrangian nature, which allows an easy treatment of
open boundaries and complex geometries, like multiplanetary systems.
%% RS

In order to study the evolution of a protoplanetary system with
several planets, we included
a second protoplanet in the simulation. 
%% WK
After the first 200~orbits of the inner planet, the inner disc has already
nearly vanished leading to a lower accretion and migration rate of the inner planet
in comparison to the outer planet. 
Due to different migration speeds of the two planets
they approach each other radially and are eventually captured
into a 2:1 mean motion resonance. 
Upon capture, the dynamical interaction between the two planets increases and
their eccentricities begin to grow considerably. 
This resonant behaviour has been found and investigated numerically
already by several authors
\citep[e.g.][]{2000MNRAS.313L..47K,2002MNRAS.333L..26N,
 2002ApJ...567..596L, kley2004}.
Depending on the subsequent evolution (distance of migration and damping),
the system might become unstable eventually.
 
Thus, the
gravitational interaction between protoplanets in the disc can explain the appearance
of resonant extra-solar planets with high eccentricities as found for example
in GJ~876, HD~82943, or 55~Cnc.

% conclusion
\section{Conclusion\label{section_conclusions}}
With the presented calculations we have modelled the dynamical evolution of
a protoplanet embedded in a disc using the Lagrangian particle method
SPH. The SPH-representation of the shear viscosity allows 
accurate treatment of the viscosity of the system, and the use of a
newly introduced artificial bulk viscosity term in the Navier-Stokes equation
inhibits mutual particle penetration. Our investigations comprised the
evolution of a single Jovian-like protoplanet in a protoplanetary accretion disc,
including the accretion of gas onto the planet and the migration of the
planet through the disc. Moreover, we examined the effects on migration
if two planets are located in the disc.\par
According to our simulations, gap formation does not
inhibit accretion onto the protoplanet. Our simulations indicate that a
protoplanet can grow up to several Jupiter masses before it has
migrated to the central star.  The
simulations with two protoplanets show a possible mechanism for the
formation of a two-planet system with a lower-mass planet near the
central star and a more massive planet with a larger semi-major axis.
The results of our SPH simulations compare favourably with those of grid
codes.
%% RS
Although the computational costs are rather high, and therefore SPH
seems to be less effective to model this problem, it is well suited to
verify simulation results achieved with other, grid-based numerical
schemes because of its completely different numerical approach.
Moreover, it may be useful for 3D-simulations or self-gravitating flows.
%% RS

\begin{acknowledgements}
Part of this work was funded by the German
Science Foundation (DFG) in the frame of the Collaborative Research
Centre SFB 382 {\em Methods and Algorithms for the Simulation of
Physical Processes on Supercomputers}. Additionally, CS 
wishes to thank the UK Astrophysical Fluid Facility (UKAFF) in
Leicester for the kind 
hospitality during a visit, where parts of this work were initiated,
and he would like to acknowledge the funding of this visit by the EU
FP5 programme.
%% RS
We are also deeply grateful for the comments and suggestions of an
anonymous referee which helped to clarify and improve the paper
considerably.  
%% RS
\end{acknowledgements}

%% RS
\appendix
\section{Numerical shear\label{appendix_numerics}}
To measure the intrinsic numerical shear viscosity of the new bulk
viscosity approach, we have performed several test simulations of the
viscously spreading ring, a model representing an idealised accretion
disc orbiting a point mass. Because an analytic solution exists for
this model (originally stated by \cite{1952ZNtrF..7a...87L} and later
by \cite{1974MNRAS.168..603L}), it is frequently used as a test
problem for numerical codes developed to simulate accretion discs
(e.g., \cite{1999JCAM...109...231S,Kley:1999:Mass}).

\begin{figure}
\resizebox{\hsize}{!}{\includegraphics{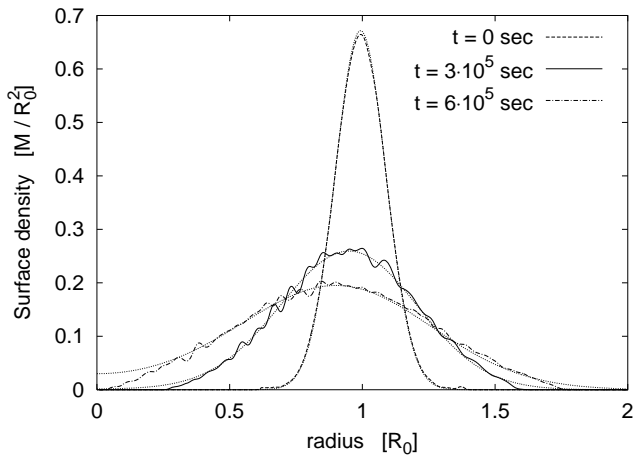}}
\caption{\label{phys-visc.ps} 
Evolution of the azimuthally averaged
surface density of the viscous ring at three different times, using
only physical viscosity in the simulation. Dotted lines denote the
analytic solutions.}
\end{figure}

Usually, modelling the physical shear viscosity by implementing the
viscous stress tensor according to Eq.~(\ref{physical_viscosity}) is
sufficient to simulate the evolution of the viscous ring. This can be
seen in Fig.~\ref{phys-visc.ps} where the results of an SPH
simulation with only physical viscosity, i.e.\ without any artificial
viscosity, is shown. Plotted is the azimuthally averaged surface
density over radius (the radial coordinate is normalized to the radius
of the initial peak, $R_0$, and the surface density is normalized to
the total mass of the accretion ring, $M/R_0^2$). For the kinematic
viscosity we chose $\nu = \nu_0 = 10^{15}$~$\mathrm{cm^2/s}$, the
value also used in the simulations of the protoplanetary disc. The
solid line represents the averaged SPH results after a simulation of
$3\times10^5$~sec, the dashed-dotted line the results after
$6\times10^5$~sec, while the dashed line indicate the initial
distribution, and the dotted lines denote the analytic solutions,
respectively. The overall time evolution of the ring is reproduced
very well. The appearing oscillations and deviations from the analytic
solutions are mainly caused by a secular viscous instability inherent
to the ring model \citep{2003AAP...399...395S}.

\begin{figure}
\resizebox{\hsize}{!}{\includegraphics{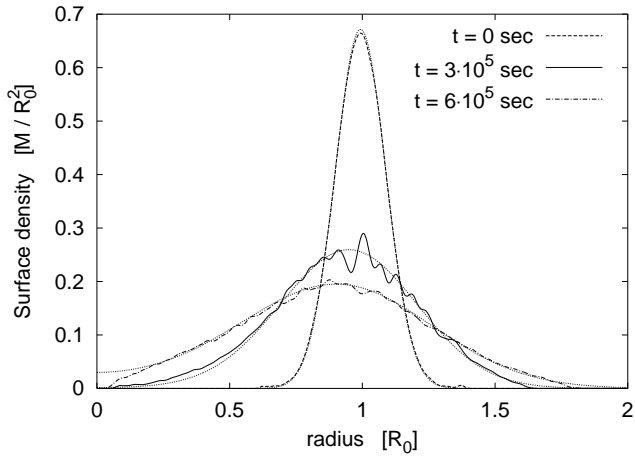}}
\caption{\label{phys+bulk-visc+XSPH.ps} 
Evolution of the azimuthally
averaged surface density of the viscous ring at three different times,
additionally using artificial bulk viscosity and XSPH together with
the physical viscosity. Dotted lines denote the analytic solutions.}
\end{figure}

However, applying only physical viscosity according to 
Eq.~(\ref{physical_viscosity}) is insufficient to simulate the dynamical
evolution of a protoplanet in a disc, because interpenetration of SPH
particles occurs such that the spiral density wave pattern in the disc
cannot be resolved. This problem is reliably prevented by the new
artificial bulk viscosity presented in Eq.~(\ref{artbulkvisceq})
in combination with the XSPH-scheme of Eq.~(\ref{xspheq}). To
test for any spurious numerical shear viscosity that may be introduced
by the usage of these algorithms, a simulation similar to that
presented in Fig.~\ref{phys-visc.ps} was performed with additional
artificial bulk viscosity, where we set $f = 0.5$ and XSPH is
switched on with $x = 0.5$. The results are shown in
Fig.~\ref{phys+bulk-visc+XSPH.ps}, where the quantities equivalent
to those in Fig.~\ref{phys-visc.ps} are plotted. As can be seen, the global
evolution of the viscous ring, which is very sensitive to the
effective shear viscosity in the simulation, seems not to be affected
at all by the new artificial viscosity approach.

\begin{figure}
\resizebox{\hsize}{!}{\includegraphics{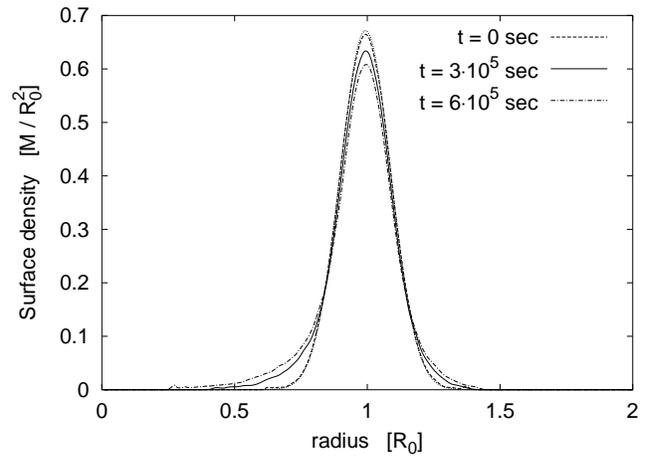}}
\caption{\label{bulk-visc.ps} 
Evolution of the azimuthally averaged
surface density of the viscous ring at three different times, applying
only artificial bulk viscosity. The dotted line represents the initial
analytic solution.}
\end{figure}

This can be tested further by a simulation where only artificial bulk
viscosity is used. The results are shown in Fig.~\ref{bulk-visc.ps}.
Any evolution of the viscous ring away from the initial distribution,
which is indicated by the dashed line, has to be due to inherent
numerical shear in the simulation. It can clearly be seen that the
intrinsic effective shear of the artificial bulk viscosity approach is
very small. A more quantitative analysis gives a value of less than
9\,\% of $\nu_0$, which can be neglected in all our simulations of the
protoplanetary disc. Note that the numerical effective shear viscosity
is not constant throughout the ring but seems to be higher at the
edges, especially at the inner edge, than in the centre.

\begin{figure}
\resizebox{\hsize}{!}{\includegraphics{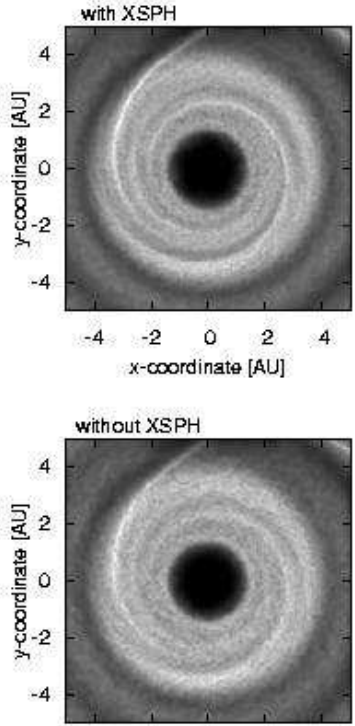}}
\caption{\label{xsph_spirals} 
Greyscale plots of the surface density
of the inner part of the protoplanetary disc in the case of the
reference model similar to Fig.~\ref{spirals1.eps} but in Cartesian
coordinates, with (upper panel) and without (lower panel) using the
XSPH-algorithm. 
%% CS
Note that the planet cannot be seen on these plots,
since it is located at a radial distance of $5.2~\mathrm{AU}$.}
\end{figure}

It has turned out that it is necessary to use the XSPH-algorithm in
addition to the artificial bulk viscosity to prevent mutual particle
penetration more effectively. This can be seen in
Fig.~\ref{xsph_spirals}, where two different simulations of the
reference model of the protoplanetary disc are shown, in the upper
panel with XSPH switched on and in the lower panel with XSPH switched
off. The surface density is plotted in greyscale, similar to the
result presented in Fig.~\ref{spirals1.eps}, but only for the inner
part of the disc and in Cartesian coordinates. Both plots show
basically the same features, but in the case with XSPH the structures
are more distinct and better resolved, while in the case without XSPH
the distribution is more diffuse. 

%% END CS

\begin{figure}
\resizebox{\hsize}{!}{\includegraphics{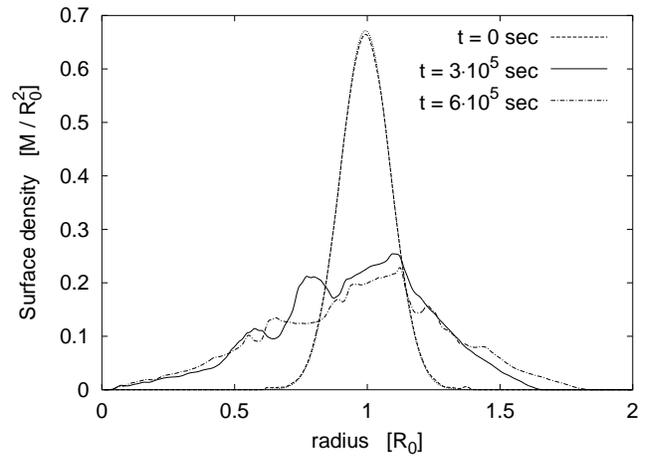}}
\caption{\label{art-visc.ps} 
Evolution of the azimuthally averaged
surface density of the viscous ring at three different times, applying
only the usual artificial viscosity. The dotted line represents the
initial analytic solution.}
\end{figure}

For the sake of completeness it should be mentioned, that a frequently used 
very efficient different approach to prevent particle interpenetration is 
the usual artificial viscosity introduced by \cite{Monaghan:1983:shock}, and
there especially the quadratic $\beta$-term. However, this artificial
viscosity suffers from a high level of spurious intrinsic shear. This
becomes obvious in the simulation of the viscous ring shown in
Fig.~\ref{art-visc.ps}, where only this artificial viscosity is used
with a rather small value of $\beta = 1$. The effective physical shear
is already of the order of $\nu_0$.

%% RS

\bibliographystyle{aa}
\bibliography{svps}
\end{document}